\documentclass[11pt,dvips]{article}
\usepackage{graphicx}
\usepackage{amsmath}
\usepackage{amssymb}
\usepackage{latexsym}
\usepackage{color}
\begin{document}
\thispagestyle{empty}
\begin{center}
\vspace{1.8cm}
 {\bf \Large{Pairwise quantum and classical correlations in  multi-qubits states via linear relative entropy}}\\
\vspace{1.5cm} {\bf M. Daoud}$^{a,b,c}${\footnote { email: {\sf
m$_{-}$daoud@hotmail.com}}}, {\bf R. Ahl Laamara}$^{d,e}$ {\footnote
{ email: {\sf ahllaamara@gmail.com}}}  and {\bf H. El Hadfi}$^{d}$
{\footnote { email: {\sf hanane.elhadfi@gmail.com}}} \\
\vspace{0.5cm}
$^{a}${\it Max Planck Institute for the Physics of Complex Systems, Dresden, Germany}\\[0.5em]
$^{b}${\it Abdus Salam  International Centre for Theoretical Physics, Trieste, Italy}\\[0.5em]
$^{c}${\it Department of Physics, Faculty of Sciences,  University Ibnou Zohr, Agadir , Morocco}\\[0.5em]
$^{d}${\it LPHE-Modeling and Simulation, Faculty  of Sciences, Rabat, Morocco}\\[0.5em]
$^{e}${\it Centre of Physics and Mathematics,  Rabat, Morocco}\\[0.5em]
\vspace{3cm} {\bf Abstract}
\end{center}
\baselineskip=18pt
\medskip

The pairwise correlations in a multi-qubit state are quantified
through a linear variant of  relative entropy. In
particular, we derive the explicit expressions of total, quantum and
classical bipartite correlations.  Two different bi-partioning
schemes are considered. We discuss the derivation of closest
product, quantum-classical and quantum-classical product states. We also investigate
the additivity relation between the various pairwise correlations existing in
pure and mixed states. As illustration, some special
cases are examined.

\newpage
\section{Introduction}

It is commonly accepted that the concept of quantum correlations
plays a central role in quantum information  processing
\cite{Horodecki-RMP-2009,Guhne}. In this context,  it is now well established that entanglement
constitutes a valuable resource for
many quantum and computational protocols,
\cite{NC-QIQC-2000,Vedral-RMP-2002}. This explains the efforts
dedicated  to the development of a general theory of quantum, classical and total
correlations
in composite quantum systems.  Various measures and several methods
were reported in the literature (for a recent review see
\cite{vedral}). Entropic based measures,  such as entanglement of
formation, linear entropy and quantum discord  have attracted considerable attention in the last decade
\cite{Rungta,Ben3,Wootters,Coffman,Vedral-et-al,Ollivier-PRL88-2001}.
Nowadays, the quantum correlations as a special feature of quantum
systems continue to be investigated from various perspectives and
for different purposes.   Recently, it has become
evident that the methods developed in  quantum information will lead to
further insight in various  areas of physics. One may quote for instance
quantum phase transitions in complex systems \cite{vedral}.

However, despite the information meaning of entropic measures, the
explicit  determination of entropy based correlations (like quantum discord) requires complex
optimization procedures, even for the simplest case
of two qubit systems. To overcome such difficulties, it was natural
to look for other measures leading to more tractable and  computable
expressions of the information contained in a multipartite quantum
system. Hence, as quantum states are represented by density
matrices,  geometric measures were used to evaluate the information
of one system about another. In this scenario, using the
Hilbert-Schmidt distance, a geometrized variant of quantum discord
was proposed in \cite{Dakic2010}. In addition, other  geometrical
approaches to quantify classical and quantum correlations in quantum
systems were discussed in the literature \cite{Bellomo1,Bellomo2}
(see also \cite{Dajka,Paula1,Aaronson,Paula2}).

Another important issue in characterizing the nature of quantum correlations in a composite system
concerns the  classification of the  total correlations  in quantum, quantum-classical and classical parts and therefore the
relationship among them. The first attempt towards a unified view of correlations in multipartite systems
was developed in \cite{Modi}  using the  notion of relative entropy. In the same vein,  a geometric formulation
based on Hilbert-Schmidt norm was reported  in \cite{Bellomo1,Bellomo2}.

The understanding of the correlation in  many body systems,
especially highly correlated ones, remains challenging  from
experimental as well as theoretical point of views. To quantify the
overall amount of  correlations in a multipartite system, it is
natural to consider pairwise correlations. In this picture, the
pairwise quantum correlations in $n$ qubit systems, possessing
exchange and parity symmetries, were considered in
\cite{DaoudPLA,Kaydi}. The original definition of quantum discord
\cite{Henderson,Ollivier} and its geometric version (geometric
quantum discord) \cite{Dakic2010} were used to derive the explicit
expressions  of pairwise correlations in the system.  However, in
the mentioned papers (by two authors of us), very little attention
has been paid to the classification of the states  according to
their degrees of quantumness.
 This issue  is especially important  if one wants
to construct a unified framework establishing the   relations among all correlations in a multi-partite system, namely classical, quantum and total correlations.
 In this respect,  we shall consider the linear relative entropy to evaluate the distance between
 a given state and its closest one without the desired property.  We develop a quantitative
 unified scheme for bipartite correlations in a multi-qubit system.   On the technical front, the linear relative entropy provides
 a significantly simple  formalism to derive computable analytical expressions of bipartite correlations.  This
is due essentially to the simplifications of  the optimization procedures which are intractable in the  relative entropy  based framework  discussed in \cite{Modi}. In the other hand,  we show that
 the quantum discord based on linear relative entropy  is equivalent  to the pairwise geometric  discord (using Hilbert-Schmidt trace) derived in  \cite{DaoudPLA} . We establish, via the linear relative entropy,
 a closed  additivity relation between the various kinds of bipartite (total, quantum and classical) correlations.  This is the second merit of
 the linear relative entropy in comparison with the geometric unified view reported in \cite{Bellomo1,Bellomo2} in which the bipartite correlations, as measured by
 Hilbert-Schmidt distance, violate the additivity relation found in \cite{Modi}.  We also derive the explicit forms of products states, the classically correlated states and the corresponding closest product states.
 This work completes further the analysis developed in  \cite{DaoudPLA,Kaydi}  concerning the characterization and the quantification of
 the various correlations existing in  $n$-qubit systems
 with parity and exchange invariance.

This paper is organized as follows. In the first section, we
introduce a specific multi-qubit state for which we shall study the
pairwise correlations (total, quantum and classical) using the
linear relative entropy. We particularly focus on $n$ qubit state possessing
parity and exchange symmetries. Two different
bi-partitions (pure and mixed) are considered.  In section 3, the notion of linear
relative entropy and its relationship with Hilbert-Schmidt norm are
presented. A closed relation involving the various kind of
correlations in a bipartite state is discussed. This relation can be viewed as a
linear variant of the additivity relation introduced in \cite{Modi}
to deal with the different correlations present in a multipartite
system in a common framework. In sections 4 and 5, the
expressions of all pairwise correlations present in the multi-qubit
system under consideration are explicitly derived (section 4
concerns the pure bipartition scheme and section 5 is devoted to mixed states).
Also, the expressions of the closest product, quantum-classical and quantum-classical
product states are obtained. To illustrate our results, some special
cases are considered in section 6. Concluding remarks close this
paper.

\section{Bipartite states in multi-qubit system and linear relative entropy}

Different methods were proposed in the literature to tackle the
issue concerning the quantification of the correlations in multi-partite systems.
For instance the Rulli-Sarandy
approach \cite{Rulli} defines the global correlation, in a
system comprising many parts,  as the maximum of correlation existing
among all possible bi-partitions. Another approach, by Z-H Ma
and coworkers \cite{Ma},  quantifies the global correlation as the sum
of all possible bi-partite correlations. Both definitions are based on the bipartite
measures of the correlations in the system. In this paper,  to
investigate the pairwise correlation in a $n$ qubit system prepared in a symmetric pure state, we
follow the procedure developed in \cite{DaoudPLA}.   The Hilbert space of this  system is
$${\cal H} =  {\cal H}_1 \otimes {\cal H}_2 \otimes \cdots \otimes {\cal H}_n, $$
where  ${\cal H}_i$ ( $i = 1,2, \cdots, n$) is spanned by the
vectors $\vert 0 \rangle $ and $\vert 1 \rangle $. An arbitrary
single  qubit writes
\begin{equation}
| \omega , \phi \rangle = \sqrt{\omega}~ |0\rangle
           + \sqrt{1- \omega} ~ e^{i \phi}~ |1\rangle, \label{coh}
\end{equation}
with  $ 0 \leq \omega \leq 1$  and $\phi \in \mathbb{R}$.  In this
work, we shall focus on the following equally weighted or balanced
multi-partite state
\begin{equation}\label{ncs}
 \vert  \omega , n  \rangle  =  {\cal N} ( \vert n, \omega, + ) + \vert n,
 \omega, -
 )),
\end{equation}
where
the states $\vert n, \omega, \pm)$ are given by
\begin{equation}\label{pss}
\vert n, \omega, \pm)  =   |  \omega , \pm  \rangle_1  \otimes |
\omega , \pm \rangle_2 \otimes \cdots \otimes  | \omega , \pm
\rangle_n,
\end{equation}
with
$$ |  \omega , +   \rangle \equiv |  \omega , 0 \rangle  \qquad |  \omega , - \rangle \equiv |  \omega , \pi  \rangle. $$
In equation (\ref{ncs}), the normalization factor ${\cal N}$  is
$$ {\cal N} = \big[ 2 + 2 s^{n} \big]^{-1/2},$$
where  $s$ denotes the scalar product between the states $|  \omega , +
\rangle $ and $|  \omega , -   \rangle $. It is defined by
\begin{equation}\label{overlap}
 s = 2\omega - 1.
\end{equation}
Hereafter, we consider only  states with $ \frac{1}{2}
\leq \omega \leq 1$ such that $ 0 \leq s \leq 1$. The situation
where $0 \leq \omega \leq  \frac{1}{2} $ can be deduced  by
replacing the state $| \omega , \phi \rangle $ by  $|1- \omega ,
\phi \rangle $. The state $| \omega , \phi \rangle $  can be
identified with the spin-$\frac{1}{2}$ coherent state with $|0\rangle
\equiv |\frac{1}{2}, -\frac{1}{2}\rangle$ and $|1\rangle \equiv
|\frac{1}{2} ,+ \frac{1}{2}\rangle$. Henceforth,  the state
(\ref{pss}) can be identified with  a spin $j = n/2$ coherent state
\cite{Kaydi}. A spin-$j$ coherent state is the tensor product of
$2j$ spin $\frac{1}{2}$ coherent states. In this respect, a pure
separable state of $n=2j$ qubits is necessary  a spin-$j$ coherent
state. Also, it is interesting to note that the states (\ref{ncs})
can be viewed as  the even  spin coherent states \cite{Kaydi}.\\

\noindent Along the line of reasoning adopted in \cite{DaoudPLA}, two different kinds of
bi-partitions of the $n$ qubit system (\ref{ncs}) are possible.  In
the first scheme,  the system is partitioned into two distinct subsystems.  In the  second
scheme, a two-qubit state  is obtained by tracing out $(n-2)$ qubit from the whole system.  These
two partitioning schemes are discussed in what follows.\\

\subsection{Pairwise partitioning and  qubit mapping}
\subsubsection{Bipartite pure states}
In the first scheme, the system is divided into two components. The first one
contains $k$  qubits $(1\le k\le n-1)$ and the second subsystem comprises the
remaining $(n-k)$ qubits. In this setting, the state (\ref{ncs}) reads as
\begin{equation}
 \vert  \omega , n  \rangle  = {\cal N} (\vert\omega,+\rangle_k \otimes \vert
\omega,+ \rangle_{n-k} +  \vert \omega,- \rangle_k \otimes \vert
\omega,- \rangle_{n-k}),\label{partition1}
\end{equation}
where
$$\vert\omega, \pm \rangle_k  =  |  \omega , \pm  \rangle_1  \otimes |
\omega , \pm \rangle_2 \otimes \cdots \otimes  | \omega , \pm
\rangle_k,
$$
$$ \vert\omega, \pm \rangle_{n-k}  =  |  \omega , \pm  \rangle_{k+1}  \otimes |
\omega , \pm \rangle_{k+2} \otimes \cdots \otimes  | \omega , \pm
\rangle_n.
$$
In order to evaluate the information of
one subsystem about another,  we introduce a pair of two logical qubits
by means of  the orthogonal basis $\{ \vert {
0} \rangle_k , \vert { 1} \rangle_k\}$ defined as
\begin{equation}
\vert { 0} \rangle_k = \frac{\vert  \omega,+\rangle_k  + \vert
\omega,-\rangle_k }{\sqrt{2(1 + s^k)}},
   \qquad \vert {
1} \rangle_k = \frac{\vert \omega,+\rangle_k - \vert
\omega,-\rangle_k }{{\sqrt{2(1- s^k)}}}\label{base1}
\end{equation}
for the  first sub-system . Similarly,  for the second subsystem,  we introduce the
following two
logical qubits $\{ \vert {0} \rangle_{n-k} , \vert {1}
\rangle_{n-k}\}$ as
\begin{equation}
\vert {0} \rangle_{n-k} = \frac{ \vert \omega,+\rangle_{n-k} +
\vert \omega, - \rangle_{n-k}}{\sqrt{2(1 + s^{n-k})}},
   \qquad
   \vert {
1} \rangle_{n-k} = \frac{\vert \omega, + \rangle_{n-k} -  \vert
\omega, - \rangle_{n-k}}{{\sqrt{2(1- s^{n-k}})}}.\label{base2}
\end{equation}
Inserting the definitions  (\ref{base1}) and (\ref{base2}) in the expression
of the multi-qubits state partitioned as in
(\ref{partition1}) and using the Schmidt decomposition,  we obtain
\begin{equation}\label{shcmidt}
\vert \omega,  n \rangle = \sqrt{\lambda_+} ~\vert {\bf 0} \rangle_k
\otimes \vert {\bf 0} \rangle_{n-k} + \sqrt{\lambda_-}~ \vert {\bf
1} \rangle_k \otimes \vert {\bf 1} \rangle_{n-k},
\end{equation}
where
the quantities given by
\begin{equation}\label{lambda+-}
\lambda_{\pm} = \frac{1}{2} \bigg(1  \pm \frac{s^k + s^{n-k}}{1 +
s^n }\bigg),
\end{equation}
are the eigenvalues of the marginal density $\rho_k $ ($\rho_k = {\rm Tr}_{(n-k)} \rho_{k,n-k}$ with  $\rho_{k,n-k} = \vert \omega,  n \rangle \langle \omega,  n \vert$) associated with the
subsystem grouping $k$ qubits. In the Schmidt decomposition  (\ref{shcmidt}), the vectors
$\vert {\bf 0} \rangle_k$ and $\vert {\bf 1} \rangle_k$ (resp. $\vert {\bf 0} \rangle_{n-k}$ and $\vert {\bf 1} \rangle_{n-k}$) are
the eigenvectors of the marginal density $\rho_k$ (resp.  $\rho_{n-k} = {\rm Tr}_{(k)} \rho_{k,n-k} $).
The
Fano-Bloch representation of  the state (\ref{shcmidt}) reads
\begin{equation}\label{rhopure}
\rho_{k,n-k} =\frac{1}{4}\bigg[\sigma_0\otimes\sigma_0+
R_{30}~\sigma_3\otimes \sigma_0+
R_{03}~\sigma_0\otimes\sigma_{3}+\sum_{i = 1}^{3}
R_{ii}~\sigma_{i}\otimes\sigma_{i}\bigg],
\end{equation}
where the correlation matrix elements are given by
\begin{equation}
 R_{30} = R_{03} =  \lambda_+ - \lambda_-, \qquad
 R_{11} =  - R_{22} = 2\sqrt{\lambda_+ \lambda_- }, \qquad R_{33} = 1.
\end{equation}
They rewrite explicitly  as
\begin{equation}\label{Rpure}
 R_{30} = R_{03} = \frac{s^k + s^{n-k}}{1 + s^n }, \quad R_{11} =
- R_{22} = \frac{\sqrt{(1- s^{2k} ) (1- s^{2(n-k)})}}{1 + s^n },
\quad R_{33} = 1
\end{equation}
 in
term of the overlapping factor $s$.

\subsubsection{ Bipartite  mixed states}
The second procedure to extract two qubit states from the entire $n$ qubit
system  consists in tracing out all qubits except two.  In this way,
one obtains $n(n-1)/2$ identical density matrices and  the reduced
density matrix describing  a two qubit subsystem writes
\begin{eqnarray}
\rho_{12} = {\cal N}^2(\vert \omega,+ , \omega,+ \rangle \langle
\omega,+, \omega,+ \vert +\vert \omega,-,\omega,-\rangle \langle
\omega,- , \omega,- | \nonumber\\+ s^{n-2} | \omega,- , \omega,-
\rangle \langle\omega,+ , \omega,+ \vert + s^{n-2} \vert \omega,+ ,
\omega,+ \rangle \langle\omega,-, \omega,- \vert ).
\end{eqnarray}
It can  be also expressed  as a two rank operator
\begin{eqnarray}\label{rho12}
\rho_{12} &=& \frac{1}{2} (1 + s^{n-2})~ \frac{{\cal N}^2}{{\cal
N}_+^2}~ \vert \omega_+ \rangle \langle \omega_+ \vert + \frac{1}{2}
(1 - s^{n-2})~ \frac{{\cal N}^2}{{\cal N}_-^2} ~\vert \omega_-
\rangle \langle \omega_- \vert,
\end{eqnarray}
where
$$\vert \omega_{\pm} \rangle = {\cal N}_{\pm} ( \vert  \omega,+ \rangle \otimes \vert  \omega,+ \rangle \pm
\vert   \omega,-  \rangle \otimes \vert   \omega,-\rangle),$$ and
$${\cal N}_{\pm}^2 = 2 \pm 2  s^{2}.$$
In the Fano-Bloch representation, the state $\rho_{12}$ (\ref{rho12}) takes the form
\begin{equation}\label{rhomixte}
\rho_{12} = \sum_{\alpha \beta} {\cal R}_{\alpha \beta}~
\sigma_{\alpha}\otimes \sigma_{\beta},
\end{equation}
where the non-vanishing matrix elements ${\cal R}_{\alpha \beta}$ $(\alpha,
\beta = 0,1,2,3)$ are
\begin{equation}\label{Rmixte}
    {\cal R}_{00} = 1, \quad {\cal R}_{11} = 2{\cal N}^2 (1- s^2), \quad {\cal R}_{22} = -2{\cal N}^2 (1-
    s^2)~s^{n-2},
{\cal R}_{33} = 2{\cal N}^2 (s^2 + s^{n-2}), \quad {\cal R}_{03} = {\cal R}_{30} = 2{\cal
N}^2 (s +  s^{n-1}).
\end{equation}

\section{Pairwise correlations using linear relative entropy}

In the unified view presented in \cite{Modi}, the relative entropy
based measures of quantum, classical and  total correlations satisfy
a closed additivity relation. Unfortunately, this approach requires
minimization procedures that are in general impossible to achieve
analytically.
 In this section, we present an alternative description to overcome these analytical problems. In this order, exploiting the pairwise qubit mapping
discussed in the previous section (see also \cite{DaoudPLA}) and the
concept of linear relative entropy, we propose a unified scheme to
deal equally with the various kinds of correlations in a bipartite
system. Hence,  the main objective of this section is to develop  a
suitable quantification of the pairwise correlations present in the
$n$-qubit state (\ref{ncs}) using  the concept of linear relative
entropy. This   provides us with a simple way to derive the explicit
expressions of  the different pairwise correlations (classical,
quantum and total) in the bipartite states $\rho_{k,n-k}$
(\ref{rhopure}) and $\rho_{12}$ (\ref{rhomixte}).

\subsection{Correlation quantifiers based on relative entropy}

To begin, we first review the essential of the unifying view of
correlations developed in \cite{Modi} based on the formalism of the  relative
entropy  defined by
\begin{equation}\label{RE-def}
S(\rho\|\sigma)=-\mathrm{Tr}(\rho\log\sigma)-S(\rho),
\end{equation}
where $S(\rho)=-\mathrm{Tr}(\rho\log\rho)$ is the von Neumann
entropy. It constitutes an adequate tool to decide about the
dissimilarity  between two quantum states. Using this measure Modi
et al \cite{Modi} have shown that the closest product state
$\pi_{\rho}$ from any density operator $\rho$ is given by the
product of the marginal matrices and the total correlation is given
by the differences of von Neumann entropies as
\begin{equation}\label{T}
T=S(\rho\|\pi_{\rho}) = S(\pi_{\rho}) - S(\rho).
\end{equation}
Similarly, quantum correlation can be viewed as a minimum loss due
to measurements in the sense of quantum mutual information. The
information in a classically correlated state is not disturbed by
local measurements. This state is called a quantum-classical state
(sometimes also referred to as classical state \cite{Modi}) and it
is of the form
\begin{equation}\label{css}
\chi_\rho=\sum_{i,j}p_{i,j}\vert{i}\rangle\langle{i}\vert\otimes\vert{j}\rangle\langle{j}\vert,
\end{equation}
where  $p_{i,j}$ are the probabilities with $\{\vert{i}\rangle,\vert{j}\rangle\}$ local basis.
 The quantum correlation is then defined as the minimal distance
between the state $\rho$ and the quantum-classical states
$\chi_\rho$. It is given by the differences between von Neumann
entropies as
\begin{equation}\label{D}
D=S(\rho\|\chi_{\rho}) =  S( \chi_{\rho}) - S(\rho).
\end{equation}
The classical correlations, defined using relative entropy, are
\begin{equation}\label{C}
C=S(\chi_{\rho}\|\pi_{\chi_{\rho}})=  S(\pi_{\chi_{\rho}})-
S(\chi_{\rho}),
\end{equation}
where $\pi_{\chi_{\rho}}$ is the nearest classical product state to
quantum-classical state $\chi_\rho$. In general, the total correlation $T$
is different from the sum of quantum and classical correlations $(T
\neq D+C)$. This difference is the relative entropy between the
closest product state $\pi_{\rho}$ and the closest product state
$\pi_{\chi_{\rho}}$ to quantum-classical state $\chi_\rho$. It is defined by
\cite{Modi}
\begin{equation}\label{L}
L=S(\pi_{\rho}\|\pi_{\chi_{\rho}})= S(\pi_{\chi_{\rho}})-
S(\pi_{\rho}),
\end{equation}
such that  the correlations as measured by relative entropy
satisfy the following remarkable additivity relation
\begin{equation}\label{additive relation RE}
 T- D- C + L = 0.
\end{equation}
As already noticed, the analytical evaluation of correlations quantifiers based
on relative entropy requires optimization techniques that are in
general very challenging. Also, the relative entropy is not
symmetrical in its arguments and subsequently  it cannot be
considered as a true distance. To get computable
expressions of classical and quantum correlations in the states
(\ref{rhopure}) and (\ref{rhomixte}), we introduce the concept of linear relative entropy.

\subsection{Correlation quantifiers based on linear relative entropy}
We define the linear relative entropy, for two arbitrary  density
matrices $\rho_1$ and $\rho_2$,  as
\begin{equation}\label{lre}
S(\rho_1 \|\rho_2) = {\rm Tr} \rho_1(\rho_1 - \rho_2),
\end{equation}
from which we
introduce the following symmetric and antisymmetric quantities
\begin{equation}\label{s+}
S_+(\rho_1 \|\rho_2)  =   S(\rho_1 \|\rho_2) +  S(\rho_2 \|\rho_1),
\end{equation}
and
\begin{equation}\label{s-}
S_-(\rho_1 \|\rho_2)  =   S(\rho_1 \|\rho_2) -  S(\rho_2 \|\rho_1).
\end{equation}
The antisymmetric part (\ref{s-}) of  the linear relative entropy
 writes as the difference of the linear entropies (or
purities) of the states $\rho_1$ and $\rho_2$. It is given by
\begin{equation}\label{s-2}
S_-(\rho_1 \|\rho_2)  =   S_2(\rho_2) -  S_2(\rho_1),
\end{equation}
in terms of the linear entropy defined by
\begin{equation}
S_2(\rho) = 1 - {\rm Tr}\rho^2.
\end{equation}
The symmetric part (\ref{s+}) rewrites as
\begin{equation}\label{s+2}
S_+(\rho_1 \|\rho_2)  =  \|\rho_1 - \rho_2\|^2 ,
\end{equation}
that is exactly the Hilbert-Schmidt distance usually used as
geometric measure of quantum correlations. We notice that the
relative entropy (\ref{RE-def}) and the linear relative entropy
(\ref{lre}) coincide only for states close to a completely  mixed
state. Indeed, in approximating   $\log \rho$ by  $  \rho -
\mathbb{I}$ where $\mathbb{I}$ the identity matrix,   the von
Neumann entropy   reduces to linear entropy and the relative entropy
is approximated by (\ref{lre}). For two-qubit states $\rho$ and
$\rho'$, expressed in Fano-Bloch representation, the Hilbert-Schmidt
distance (\ref{s+2}) is
\begin{eqnarray}\label{distance-def}
 S_+(\rho \|\rho') & = & \frac{1}{4} \sum_{\alpha,\beta} ~ (R_{\alpha\beta} -
 R'_{\alpha\beta})^2,
\end{eqnarray}
where $\alpha, \beta = 0, 1, 2, 3$. The linear variant of
total correlation $T$ (\ref{T}), quantum correlation $D$ (\ref{D}),
classical correlation $C$ (\ref{C}) and the quantity $L$ (\ref{L})
are  respectively defined  by
\begin{eqnarray}\label{T2}
T_2 =  S_-(\rho \| \pi_\rho),\qquad D_2 =  S_-(\rho \| \chi_\rho),
\qquad C_2 = S_-(\chi_\rho\| \pi_{\chi_\rho}), \qquad L_2 =
S_-(\pi_\rho\| \pi_{\chi_\rho}).
\end{eqnarray}
Using (\ref{s-2}), it is easy to check that  they satisfy the additivity relation
\begin{eqnarray}\label{additivity 2}
T_2 - D_2 -C_2 + L_2 = 0
\end{eqnarray}
to be compared with  (\ref{additive relation RE}). In the next section, we shall derive
the analytical expressions of the different
correlations in the states (\ref{rhopure}) and (\ref{rhomixte}).

\section{ Correlations in  bipartite  pure states}

Having introduced the linear relative entropy (\ref{s-2}), we now
determine the pairwise correlations in the states of type
(\ref{rhopure}) corresponding to the pure partitioning scheme of the
$n$-qubit state (\ref{ncs}). A special attention is also devoted to
the explicit forms of its closest product, quantum-classical and quantum-classical
product states.

\subsection{Total correlation}\label{subsect-prho}
To evaluate the explicit form of total correlation $T_2$ (\ref{T2}), we first
determine the closest product state from  the pure state (\ref{rhopure}). An arbitrary product state $ \pi_{\rho_{k,n-k}} = \varrho_k \otimes \varrho_{n-k} $  writes
\begin{eqnarray}
  \pi_{\rho_{k,n-k}} & = & \frac{1}{4}\bigg[ \sigma_{0}\otimes \sigma_{0} +\sum_{i}^{3}(a_{i}~\sigma_{i}\otimes \sigma_{0}
   +b_{i} ~\sigma_{0}\otimes\sigma_{i})+\sum_{i,j=1}^{3} a_{i}b_{j} ~
   \sigma_{i}\otimes\sigma_{j}\bigg],
\end{eqnarray}
where $(a_1,a_2,a_3)$ and $(b_1,b_2,b_3)$ stand for the  Bloch
vectors of the operators $\varrho_k$ and $\varrho_{n-k}$ respectively. We
note that the density matrix $\rho_{k,n-k}$ (\ref{rhopure}) is
invariant under exchange  and parity symmetries. This implies that
the closest product state must be also invariant under these two
operations. This simplify considerably the minimization of the
distance, defined by (\ref{distance-def}), between $\rho_{k,n-k}$ and $\pi_{\rho_{k,n-k}}$
. Indeed, the exchange symmetry of the state
$\pi_{\rho_{k,n-k}}$ imposes
$$ a_i = b_i, \qquad i = 1,2,3.$$
Also, the parity invariance of the density matrix $([ \pi_{\rho_{k,n-k}} , \sigma_3 \otimes \sigma_3] = 0)$ gives
$$ a_i = b_i = 0, \qquad i = 1,2.$$
It follows that the distance (\ref{distance-def}), separating the
states $\rho_{k,n-k}$  and $\pi_{\rho_{k,n-k}}$,   writes
\begin{eqnarray}
S_+(\rho_{k,n-k} \|\pi_{\rho_{k,n-k}}) & = & \frac{1}{4}
\big[2(R_{30} - a_3)^2  + R_{11}^2 + R_{22}^2 +  (R_{33} -
a_3^2)^2\big].
\end{eqnarray}
It is simple to check  that the minimal distance is obtained for
\begin{eqnarray}\label{solutiona3-pure}
a_3^3  = R_{30} \quad \Longleftrightarrow  \quad a_3 =  \sqrt[3]{
\lambda_+ - \lambda_-}.
\end{eqnarray}
Explicitly, we have
\begin{eqnarray}\label{a3}
a_3 = \sqrt[3]{\frac{s^k + s^{n-k}}{1+s^n}},
\end{eqnarray}
in terms
of the overlap parameter $s$. Thus, the nearest product state to
the pure state $\rho_{k,n-k}$ is
\begin{eqnarray}\label{pirho}
\pi_{\rho_{k,n-k}} & = & \frac{1}{4}\bigg[ \sigma_{0}\otimes
\sigma_{0} + a_{3}~\sigma_{3}\otimes \sigma_{0}
   +a_{3} ~\sigma_{0}\otimes\sigma_{3} +  a^2_{3} ~
   \sigma_{3}\otimes\sigma_{3}\bigg],
\end{eqnarray}
and  the total correlation is given by
\begin{equation}\label{T2pure}
T_2 (\rho_{k,n-k}) = \frac{1}{4}\big[ 2 ( R^2_{03} - a^2_3 ) +
R^2_{11} + R_{22}^2 +  (R^2_{33}-a^4_3)],
\end{equation}
where the correlation matrix elements and the quantity $a_3$  are
respectively given by  (\ref{Rpure}) and (\ref{a3}).

\subsection{Quantum correlation}
When the measurement is taken on the subsystem containing $k$
qubits, the quantum-classical state $\chi_{\rho_{k,n-k}}$ has the form
({\ref{css}}) \cite{Dakic2010}. In this case, one can show that  $
{\rm Tr} \rho\chi = {\rm Tr} \chi^2$.  This implies that the quantum
correlation defined in (\ref{T2}) coincides with the Hilbert-Schmidt
distance between the states $\rho_{k,n-k}$ and
$\chi_{\rho_{k,n-k}}$. In other words, the quantum correlation $D_2$ based on linear
relative entropy
coincides with the  geometric quantum discord  introduced in
\cite{Dakic2010}. Subsequently, following the standard method to
derive the geometric measure of quantum  discord for a two-qubit
system  \cite{Dakic2010}, one gets
\begin{equation}
D_2 =\frac{1}{4}\left(\lambda_1 + \lambda_2 + \lambda_3
-\lambda_{\rm {max}}\right),
\end{equation}
where $\lambda_{\rm{max}} = {\rm max} (\lambda_1 , \lambda_2 , \lambda_3)$ is the largest eigenvalue of  the matrix $K$ defined by
\begin{equation}
K := xx^{T}+RR^{T}, \label{matrix K}
\end{equation}
where $x=(x_{1} = R_{03},x_{2} = 0 ,x_{3} = 0)^{T}$, $R$ is the
matrix with elements $R_{ij}$ $(i,j=1,2,3)$. For the bipartite  pure state
(\ref{rhopure}),  the eigenvalues of the matrix $K$ are
$$ \lambda_1 =   \lambda_2 =  4 \lambda_+\lambda_-  \qquad \lambda_3 =  2(\lambda^2_+ + \lambda^2_- ) $$
where $\lambda_{\pm}$ are given by (\ref{lambda+-}). Clearly, $\lambda_3$ is the largest eigenvalue of the matrix $K$ and the quantum correlation takes the simple form
\begin{equation}\label{D2pure}
D_2 (\rho_{k,n-k})  =\frac{1}{4}\left(\lambda_1 + \lambda_2 \right)
= 2 \lambda_+\lambda_-.
\end{equation}
To find the explicit expression of the closest quantum-classical
state to $\rho_{k,n-k}$, we follow the
 procedure reported in \cite{Dakic2010} for an arbitrary two qubit system.  By a straightforward calculation, one obtains
\begin{equation}\label{cspure}
   \chi_{\rho_{k,n-k}} =\frac{1}{4}[\sigma_0\otimes\sigma_0+R_{30}~\sigma_3\otimes
\sigma_0+
R_{30}~\sigma_0\otimes\sigma_{3}+R_{33}~\sigma_{3}\otimes\sigma_{3}].
\end{equation}
\subsection{ Classical correlations }
The quantum-classical state  $\chi_{\rho_{k,n-k}}$ (\ref{cspure}) possesses
parity and exchange symmetries. Thus, its closest product  can be
derived by reproducing the method used to find
 the product state  $\pi_{\rho_{k,n-k}}$ given
by (\ref{pirho}). This yields
\begin{equation}\label{productcspure}
\pi_{{\chi_{\rho_{k,n-k}}}} =  \frac{1}{4}\bigg[ \sigma_{0}\otimes
\sigma_{0} + a_{3}~\sigma_{3}\otimes \sigma_{0}
   +a_{3} ~\sigma_{0}\otimes\sigma_{3} +  a^2_{3}  ~
   \sigma_{3}\otimes\sigma_{3}\bigg],
\end{equation}
which coincides with $\pi_{\rho_{k,n-k}}$ (\ref{pirho}). Reporting the expressions of quantum-classical state
$\chi_{\rho_{k,n-k}}$ (\ref{cspure})  and its closest product state
$\pi_{{\chi_{\rho_{k,n-k}}}}$ (\ref{productcspure}) in
(\ref{T2}), the classical correlation
takes the form
\begin{eqnarray}\label{C2pure}
C_2 (\rho_{k,n-k}) =\frac{1}{4}\big[ 2 ( R^2_{03} - a^2_3 )  + (R^2_{33}-a^4_3)].
\end{eqnarray}
Finally, since the product of the quantum-classical state
$\pi_{{\chi_{\rho_{k,n-k}}}}$ coincides with the product state
$\pi_{\rho_{k,n-k}}$, one has
\begin{eqnarray}\label{L2pure}
L_2(\rho_{k,n-k}) = 0,
\end{eqnarray}
and using the results (\ref{T2pure}), (\ref{D2pure}), (\ref{C2pure}) and (\ref{L2pure}),  one shows that  the additivity relation
(\ref{additivity 2}) is satisfied as expected.

\section{ Correlations in bipartite  mixed states}

We now come to the second partitioning scheme. In order to evaluate the total correlation in the state
(\ref{rhomixte}), we determine its closest state product. The explicit expressions
of the nearest quantum-classical states to the state (\ref{rhomixte}) are needed to derive
the quantum correlation contained in the state  (\ref{rhomixte}). Finally, the computation of the classical correlation
in (\ref{rhomixte}) requires
the explicit forms of the closest product states to
the quantum-classical state.

\subsection{Total correlation}\label{subsect-prho}
To obtain the nearest product state from the two-qubit state
$\rho_{12}$   (\ref{rhomixte}), we write the product state as $
\pi_{\rho_{12}} = \rho_1 \otimes \rho_2 $. We note that,  since the
bipartite state $\rho_{12}$ is invariant by exchanging the roles of
the qubits 1 and 2, the reduced operators $\rho_1$ and $\rho_2$
must be identical. Furthermore,  the
density matrix (\ref{rhomixte}) commutes with the operator $\sigma_3
\otimes \sigma_3$ (parity invariance). It follows that the product
state $\pi_{\rho_{12}}  $ is necessarily of the form
\begin{eqnarray}\label{pirho12}
  \pi_{\rho_{12}}  =  \rho_1 \otimes  \rho_2  = \frac{1}{4}\bigg[ \sigma_{0}\otimes \sigma_{0} + c_{3}~\sigma_{3}\otimes \sigma_{0}
   +c_{3}~ \sigma_{0}\otimes\sigma_{3} + c^2_3
   ~\sigma_{3}\otimes\sigma_{3}\bigg].
\end{eqnarray}
where $c_3 \in \mathbb{R}$. Using  (\ref{s+}),  the distance between the states $\rho_{12}$ and
$\pi_{\rho_{12}}$ is
\begin{eqnarray}
S_+(\rho_{12}, \pi_{\rho_{12}}) & = & \frac{1}{4} \big[2({\cal R}_{30} -
c_3)^2  + {\cal R}_{11}^2 + {\cal R}_{22}^2 + ({\cal R}_{33} - c_3^2)^2\big],
\end{eqnarray}
which is a function of the  parameter $c_3$. The
minimal distance is obtained by setting at zero the derivative of the
distance $S_+(\rho_{12}, \pi_{\rho_{12}})$. This gives the following
 equation
\begin{eqnarray}
c_3^3  + c_3 (1 - {\cal R}_{33}) - {\cal R}_{30}  = 0.
\end{eqnarray}
By Cardano's formula,  the unique real root of  this  cubic equation
is
\begin{eqnarray}\label{solutiona3-mix}
c_3 = \sqrt[3]{ \frac{  \sqrt{\Delta} + {\cal R}_{30}}{2}} -
\sqrt[3]{\frac{ \sqrt{\Delta} - {\cal R}_{30}}{2}},
\end{eqnarray}
where
$$\Delta = {\cal R}_{30}^2 + \frac{4}{27}(1 - {\cal R}_{33})^3,$$
is non negative ( ${\cal R}_{33} \leq 1$). Combining the definitions  (\ref{s-})
 and (\ref{T2}), the total correlation in the state
(\ref{rhomixte}) , as measured by linear relative entropy, writes
\begin{equation}\label{T2c1c2}
T_2(\rho_{12}) = \frac{1}{4}\big[ 2 ( {\cal R}^2_{03} - c^2_3 ) + {\cal R}^2_{11}
+ {\cal R}^2_{22} + ({\cal R}^2_{33}-c^4_3)],
\end{equation}
where the correlation matrix elements are given by (\ref{Rmixte})
and the quantity $c_3$  (\ref{solutiona3-mix}) can be
re-equated in terms of the overlapping parameter $s$ so that the
total correlation is a function of the variable $s$.
\subsection{Quantum correlation}
For a measurement  performed on the first qubit of
$\rho_{12}$ (\ref{rhomixte}), the general form of quantum-classical state is
given by (\ref{css}). As in the pure partitioning
scheme, one verifies that the quantum correlation $D_2(\rho_{12})$
coincides with the geometric measure of quantum discord
$D_\mathrm{g}(\rho_{12})$ defined as the  Hilbert-Schmidt distance between the state
$\rho_{12}$ and its closest quantum-classical states. Hence, using the general method
proposed in \cite{Dakic2010}
(see also \cite{DaoudPLA} for more details), it is easy to obtain
 \begin{equation}
 D_2(\rho_{12}) \equiv D_\mathrm{g}(\rho_{12}) = \frac{1}{4}~ {\rm min}\{ \lambda_1 + \lambda_2 ~,~ \lambda_2 +
 \lambda_3\},
\end{equation}
where
\begin{equation}\label{lambda1}
\lambda_1 = {\cal R}_{11}^2 = 4 {\cal N}^4 (1 - s^2)^2,
\end{equation}
\begin{equation}\label{lambda2}
\lambda_2 = {\cal R}_{22}^2 = 4 {\cal N}^4 s^{2(n-2)}(1 - s^2)^2,
\end{equation}
\begin{equation}\label{lambda3}
\lambda_3 = {\cal R}_{03}^2 + {\cal R}_{33}^2 = 4 {\cal N}^4 \bigg[(1 + s^2)(s^2 +
s^{2(n-2)}) + 4s^n \bigg].
\end{equation}
To write the explicit expression of the closest quantum-classical state to
$\rho_{12}$, we exploit the optimization procedure developed in
\cite{Dakic2010}. For this end, we discuss separately the situations
where $\lambda_1 \leq \lambda_3 $ and $\lambda_3 \leq \lambda_1$.
The condition $\lambda_1 \leq \lambda_3 $ writes as
\begin{equation}
(s^2 + 1)(1+s^{n-2}) - 2(1 - s^2) \ge 0.\label{conditionplus}
\end{equation}
In this case, the closest quantum-classical state is obtained as
\begin{equation}\label{classical state first case}
   \chi_{\rho_{12}}^{-}=\frac{1}{4}[\sigma_0\otimes\sigma_0+{\cal R}_{30}\sigma_3\otimes
\sigma_0+
{\cal R}_{30}\sigma_0\otimes\sigma_{3}+{\cal R}_{33}\sigma_{3}\otimes\sigma_{3}],
\end{equation}
where the subscript $^{-}$ refers  to the situation where the
difference $\lambda_1- \lambda_3$ is non positive. It follows that
the  quantum correlation present in the two qubit state $\rho_{12}$
is
\begin{eqnarray}
 D^-_2(\rho_{12}) = S_-(\rho_{12} \| \chi_{\rho_{12}}^{-})  =   \frac{1}{4}(\lambda_1 + \lambda_2)  =  \frac{1}{4} ( {\cal R}_{11}^2  +
 {\cal R}_{22}^2),
\end{eqnarray}
which rewrites as
\begin{equation}
 D^-_2(\rho_{12})= D^-_\mathrm{g}(\rho_{12}) = \frac{1}{4}  \frac{(1+
s^{2(n-2)})(1 - s^2)^2}{(1 + s^n )^2}.\label{Dgplus}
\end{equation}
The second class of closest quantum-classical states is obtained in the case
where $\lambda_1 - \lambda_3 \geq 0$ corresponding  to the situation
when the condition
\begin{equation}
(s^2 + 1)(1+s^{n-2}) - 2(1 - s^2) \le 0 \label{conditionmoins}
\end{equation}
is satisfied.  In this case,  the closest quantum-classical states  take the
form
\begin{eqnarray}\label{classical state second case}
\chi_{\rho_{12}}^{+}&=&\frac{1}{4}\Big[
\sigma_0\otimes\sigma_0+{\cal R}_{03}\sigma_0\otimes\sigma_{3}
   + {\cal R}_{11}\sigma_1\otimes\sigma_1 \Big],
\end{eqnarray}
where now the subscript $^{+}$ refers to the sign of the difference $\lambda_1-
\lambda_3$. It follows that the quantum correlation is
\begin{eqnarray}
  D^+_2(\rho_{12}) = S_-(\rho \| \chi_{\rho}^{+})  =  \frac{1}{4}(\lambda_2 + \lambda_3)  =  \frac{1}{4} ( {\cal R}_{22}^2  + {\cal R}_{03}^2 +
  {\cal R}_{33}^2).
\end{eqnarray}
Explicitly, it is given by
\begin{equation}
D^+_2(\rho_{12}) = D^+_\mathrm{g}(\rho_{12}) = \frac{1}{4}
\frac{(s^2 + s^{2(n-2)})(1+s^2) + 4s^n + s^{2(n-2)}(1 - s^2)^2}{(1 +
s^n )^2}.\label{Dgmoins}
\end{equation}
From equations (\ref{classical state first case}) and
(\ref{classical state second case}), it is simple to check that
 the closest quantum-classical states $\chi_{\rho_{12}}^{\pm}$  satisfy the
relation
$$ {\rm Tr} \rho_{12} \chi_{\rho_{12}}^{\pm}  = {\rm Tr}  {(\chi_{\rho_{12}}^{\pm})}^2$$
which implies that the geometric discord $D_\mathrm{g}(\rho_{12})$ ,
as measured by Hilbert-Schmidt norm, coincides  indeed with quantum
correlation $D_2(\rho_{12})$ defined by means of linear relative
entropy.

\subsection{ Classical correlations }

To evaluate the classical correlation defined by (\ref{T2}), one has
to determine first the closest product states to quantum-classical states
$\chi_{\rho_{12}}^{\pm}$  given by (\ref{classical state first
case}) and (\ref{classical state second case}). Let us denote them
by $\pi_{\chi_{\rho_{12}}^{\pm}}$. To derive  the product state
$\pi_{\chi_{\rho_{12}}^{-}}$, we notice that the
permutation and parity symmetries leave the state
$\chi_{\rho_{12}}^{-}$ (\ref{classical state first case}) invariant.
Therefore, reiterating the method discussed above in obtaining the
product states ($\pi_{\rho_{12}}$ for instance), it is simple to check
that the closest product state to quantum-classical state
$\chi_{\rho_{12}}^{-}$ (\ref{classical state first case}) is
\begin{eqnarray}\label{pichi-}
\pi_{{\chi^{-}_{\rho_{12}}}} =  \frac{1}{4}\bigg[ \sigma_{0}\otimes
\sigma_{0} + c_{3}\sigma_{3}\otimes \sigma_{0}
   +c_{3} \sigma_{0}\otimes\sigma_{3} +  c^2_{3}
   \sigma_{3}\otimes\sigma_{3}\bigg]
\end{eqnarray}
which coincides with the product state  $\pi_{\rho_{12}}$.  The
classical correlation (\ref{T2}) is then given by
\begin{eqnarray}\label{c2-}
C^-_2(\rho_{12}) =\frac{1}{4}\big[ 2 ( {\cal R}^2_{03} - c^2_3 )  + ({\cal R}^2_{33}-c^4_3)].
\end{eqnarray}
For the state $\chi_{\rho_{12}}^{+}$, the situation is slightly
different. This state is not invariant under the permutation of the
qubits 1 and 2. Only  the parity symmetry is preserved. The general
form of the closest classical product states, symmetric under parity
transformation, is
\begin{eqnarray}
\pi^+_{\chi_{\rho_{12}}} & = & \frac{1}{4}\bigg[ \sigma_{0}\otimes \sigma_{0} +  \alpha_3\sigma_{3}\otimes \sigma_{0}
   +  \beta_3 \sigma_{0}\otimes\sigma_{3} +  \alpha_3 \beta_3 \sigma_{3}\otimes\sigma_{3}\bigg],
\end{eqnarray}
where  $\alpha_3$ and $\beta_3$ stand for the variables to be
optimized to get the minimal Hilbert-Schmidt distance  between the
states $\chi_{\rho_{12}}^{+}$ and $\pi_{\chi^+_{\rho_{12}}}$. A
simple calculation leads to
$$ \alpha_3 = 0, \qquad \beta_3 = {\cal R}_{03}.$$
Then,  the closest product state writes
\begin{equation}\label{pichi+}
\pi_{\chi^+_{\rho_{12}}} =\frac{1}{4}[
\sigma_0\otimes\sigma_0+{\cal R}_{03}\sigma_0\otimes\sigma_{3}],
\end{equation}
and the  classical correlation reads
\begin{eqnarray}\label{c2+}
C^+_2(\rho_{12})  =\frac{1}{4} {\cal R}^2_{11}.
\end{eqnarray}
Using the expression of the product state  $\pi_{\rho_{12}}$ (\ref
{pirho12}) and the classical product states
$\pi_{{\chi^{-}_{\rho_{12}}}}$  (\ref{pichi-}) and
$\pi_{{\chi^{+}_{\rho_{12}}}}$ (\ref{pichi+}), the expression of the
quantity  $L_2$ (\ref{T2}) is given by
\begin{eqnarray} \label{L2+-}
L^-_2 (\rho_{12}) = 0, \qquad           L^+_2(\rho_{12}) =
\frac{1}{4}\big[ 2 c^2_3   +  c^4_3 - {\cal R}^2_{03}],
\end{eqnarray}
which satisfy
\begin{eqnarray} \label{clos-add-rel}
 L^{\pm}_2 = D^{\pm}_2 + C^{\pm}_2 - T_2,
\end{eqnarray}
and one has the additivity relation (\ref{additivity 2}).  The analysis presented in this section
show clearly the advantages of the linear relative entropy (\ref{lre}) in deriving  the
 expressions of the different correlations explicitly  in terms of
the overlapping parameter $s$ and the number $n$ of qubits in the state (\ref{ncs}). This is essentially due
to the simplification arising in determining
the product, quantum-classical and quantum-classical product states
for the bipartite states $\rho_{12}$.   To analyze the behavior
of the obtained correlations as functions of $n$ and the parameter $s$, we shall discuss some specific cases.

\section{ Illustration: some special cases}
As illustration, we consider some special instances of the states (\ref{ncs}). We
shall especially give the analytical expressions of the total, quantum and classical pairwise
correlations for $n = 2$ and
$n = 3$. Numerical results for states containing more qubits are
also discussed.
\subsection{ Two qubits}
For $n=2$,  the total correlation (\ref{T2pure}) writes
\begin{equation}\label{T2puren=2}
T_2 (\rho_{1,1}) = \frac{1}{4}\Bigg[ \bigg( 1 - 2
\bigg(\frac{2s}{1+s^2}\bigg)^{\frac{2}{3}}\bigg) \bigg( 1 -
\bigg(\frac{2s}{1+s^2}\bigg)^{\frac{4}{3}}\bigg) +2 \bigg(
\frac{1-s^2}{1+s^2} \bigg)^2 \Bigg].
\end{equation}
From equation (\ref{D2pure}), the quantum correlation reads
\begin{equation}\label{D2puren=2}
D_2 (\rho_{1,1}) = \frac{1}{2} \bigg( \frac{1-s^2}{1+s^2} \bigg)^2,
\end{equation}
and using (\ref{C2pure}), the classical correlation is given by
\begin{equation}\label{C2puren=2}
C_2 (\rho_{1,1}) = \frac{1}{4} \bigg( 1 - 2
\bigg(\frac{2s}{1+s^2}\bigg)^{\frac{2}{3}}\bigg) \bigg( 1 -
\bigg(\frac{2s}{1+s^2}\bigg)^{\frac{4}{3}}\bigg).
\end{equation}
Clearly, we have the additivity relation
\begin{equation}\label{T2+D2+C2puren=2}
T_2 (\rho_{1,1}) = D_2 (\rho_{1,1})+ C_2 (\rho_{1,1}).
\end{equation}
The quantity $L_2$ (\ref{L2pure}) is zero for pure states.
These results can be equivalently  obtained from ones derived
for the second bipartite partitioning scheme. Indeed, for $n=2$  the
state $\rho_{12}$ (\ref{rho12}) is pure. Also, it is simple to
verify that for $n=2$ the solutions (\ref{solutiona3-pure}) and
(\ref{solutiona3-mix}) are identical. It follows that,  for $n=2$,
total correlation $T_2 (\rho_{1,1})$ and $T_2 (\rho_{12})$ coincide.
Moreover,  the condition (\ref{conditionplus}) is always satisfied
for $n=2$. This implies that the quantum correlation is given by
\begin{equation}
D_2 (\rho_{12})\equiv D^-_2 (\rho_{1,1}) = \frac{1}{2} \bigg(
\frac{1-s^2}{1+s^2} \bigg)^2,
\end{equation}
which is exactly the quantum discord (\ref{D2puren=2}). Similarly,
using the fact that the solutions (\ref{solutiona3-pure}) and
(\ref{solutiona3-mix}) are identical for $n=2$, one has
$$
 C_2 (\rho_{12}) =  C_2 (\rho_{1,1}).
$$
\subsection{ Three qubits}
The situation is slightly different for $n = 3$. We shall focus only on the mixed partitioning scheme. In this case, the
correlation matrix elements (\ref{Rmixte}) read
\begin{equation}\label{Rn=3}
    {\cal R}_{00} = 1, \quad {\cal R}_{11} = \frac{1- s^2}{1+s^3}, \quad {\cal R}_{22} = -s\frac{1- s^2}{1+s^3},
{\cal R}_{33} = \frac{s+ s^2}{1+s^3}, \quad {\cal R}_{03} = {\cal R}_{30} = \frac{s+
s^2}{1+s^3}.
\end{equation}
Reporting $ {\cal R}_{03}$ and $ {\cal R}_{33}$ in (\ref{solutiona3-mix}), one has
the expression of $c_3$ (\ref{solutiona3-mix}) and subsequently the total correlation (\ref{T2c1c2}) writes as
a function of the overlap $s$. The behavior of $T_2$
versus $s$ is given in figure 1. The condition (\ref{conditionplus}) (resp. (\ref{conditionmoins}))
is satisfied when $ \sqrt{2} - 1 \leq s \leq 1$ (resp. $ 0 \leq
s\leq \sqrt{2} - 1$). It follows that the quantum discord is given
by
$$  D^-_2 (\rho_{12}) =  \frac{1}{4} \frac{(1-s^2)^2(1+s^2)}{(1+s^3)^2} $$
for $ \sqrt{2} - 1 \leq s \leq 1$, and
$$ D^+_2 (\rho_{12}) =  \frac{1}{4} \frac{s^2(1+s)^2(2 + (1-s)^2)}{(1+s^3)^2} $$
for $ 0 \leq s \leq \sqrt{2} - 1$. Furthermore, the classical correlation $C^-_2 (\rho_{12})$ (resp.
$C^+_2 (\rho_{12})$) when $ \sqrt{2} - 1 \leq s \leq 1$ (resp. $0
\leq s\leq \sqrt{2} - 1$ ) is obtainable from the equation
(\ref{c2-}) (resp.\ref{c2+}) by simply replacing the correlation
matrix elements and $c_3$ by their corresponding expressions in
terms of the parameter $s$ for $n=3$. Similarly, one can write the
explicit form of the quantity  $L^+_2 (\rho_{12})$ (\ref{L2+-}) as a
function of $s$. The total pairwise correlations $T_2$ (\ref{T2c1c2}) for the states
 comprising three qubits or more $(n \geq 3)$  are depicted in
figure 1. It is remarkable that the maximal amount of total
correlation, as measured by linear relative entropy, exists in
states comprising three qubits. It is interesting also to note
that the  total pairwise correlation in the states $\rho_{12}$
(\ref{rho12}) becomes identical for all the states containing four
qubits and more.
\begin{center}
\includegraphics[width=4in]{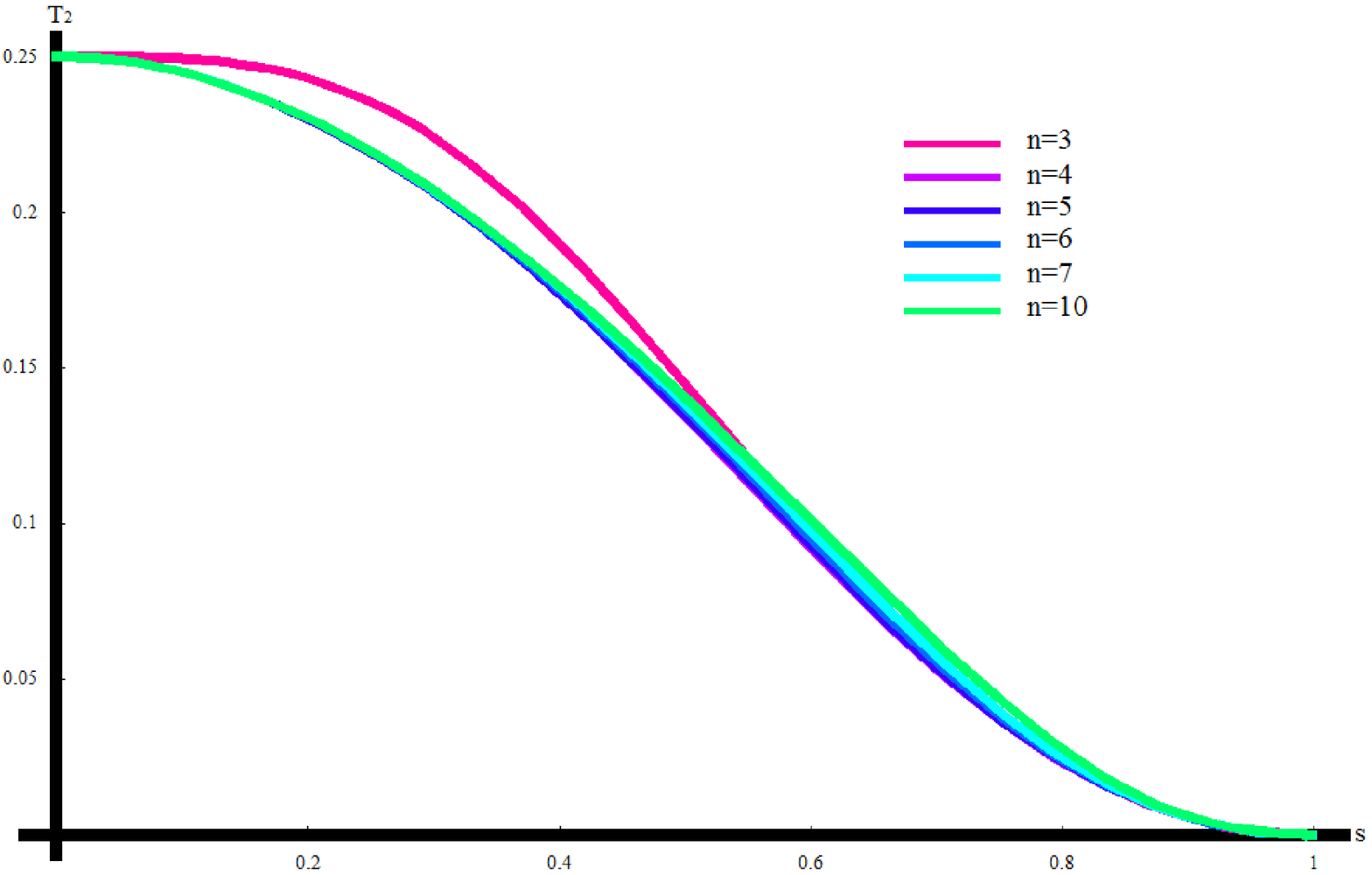}\\
{\bf Figure 1.}  {\sf The total pairwise correlation  $T_2$ versus
the overlapping $s$ for different values of $n$.}
\end{center}
We also consider the behavior of quantum
discord given by the equations (\ref{Dgplus}) and (\ref{Dgmoins})
for different values of qubits number $n$. This is reported
in figure 2. The quantum discord increases first and  start
suddenly decreasing to vanishes for $s = 1$. The sudden change of
quantum discord is responsible of the violation of the additivity
relation and induces the discontinuity of
classical correlations as it is shown in figure 3. It is also
clearly seen that the pairwise quantum correlations have
approximately the same behavior for states with $n \geq 4 $. This is
similar to what happens with total correlation (see figure 1). The
figure 3 presenting the variation of classical correlation versus
the overlap parameter  $s$ shows two important features. First, like
total and quantum correlations in states with $n \geq 4$, the
classical correlation becomes independent of $n$. The second
important feature is that the pairwise classical correlation, as
measured by linear relative entropy, is discontinuous when crossing
the point where the quantum discord suddenly changes (see figure 2). This reflects the fact that the deviation from the additivity relation (
$T_2 \neq D^+_2 + C^+_2$) and the discontinuity of classical
correlation are  deeply related to the sudden change of quantum
discord. To study the violation of the additivity relation, we
reported in figure 4, the difference $L_2$ between total correlation
$T_2$ and the sum of quantum and classical correlation $(D_2+C_2)$.
For states with $ n \geq 5$, the behavior of  $L_2$ versus $s$
becomes slightly identical when $n$ increases (figure 4). The
violation of the additivity relation is more  prominent  for states
containing small number of qubits.

\begin{center}
\includegraphics[width=4in]{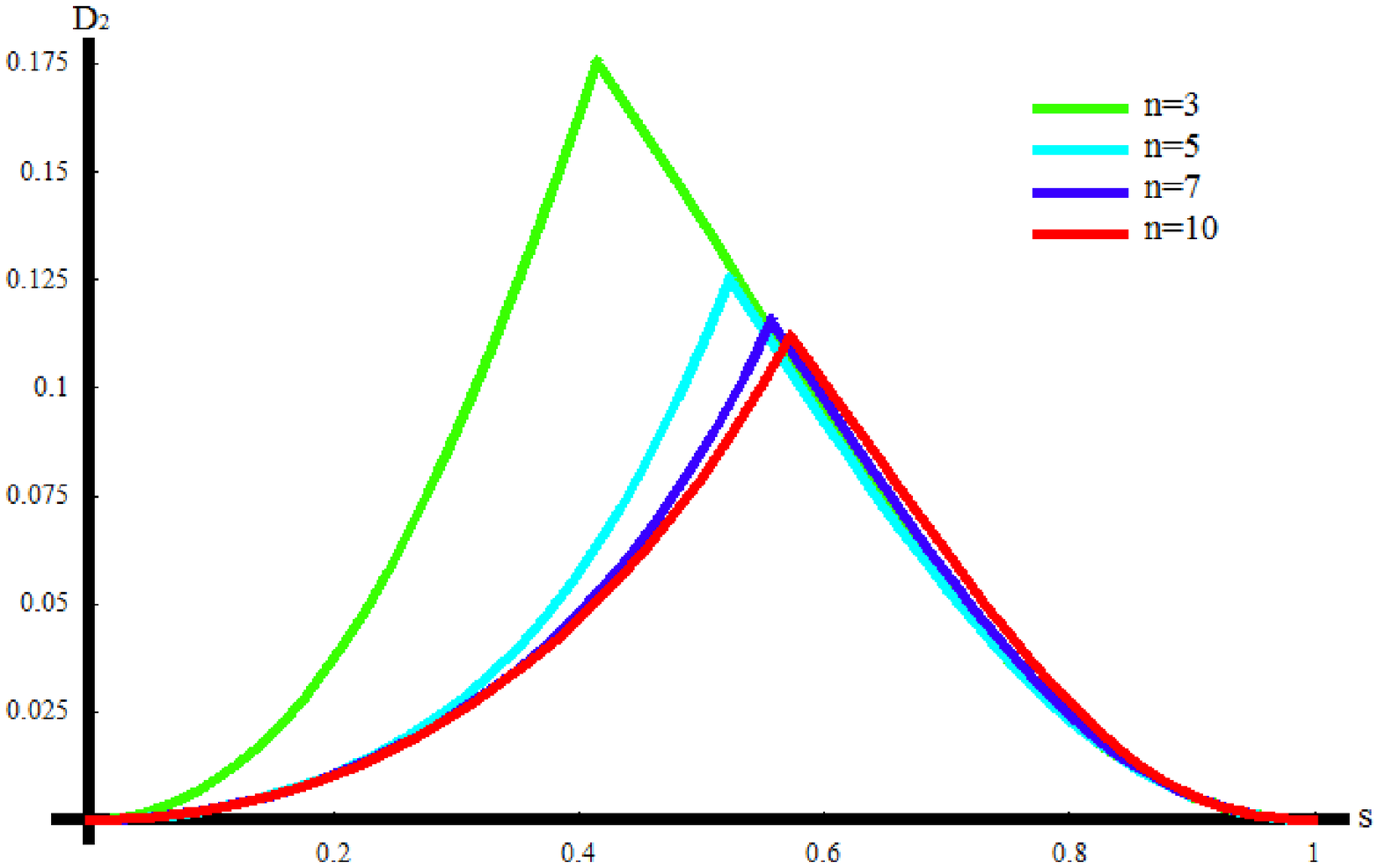}\\
{\bf Figure 2.}  {\sf The quantum discord $D_2$ versus the
overlapping $s$ for different values of $n$.}
\end{center}

\begin{center}
\includegraphics[width=4in]{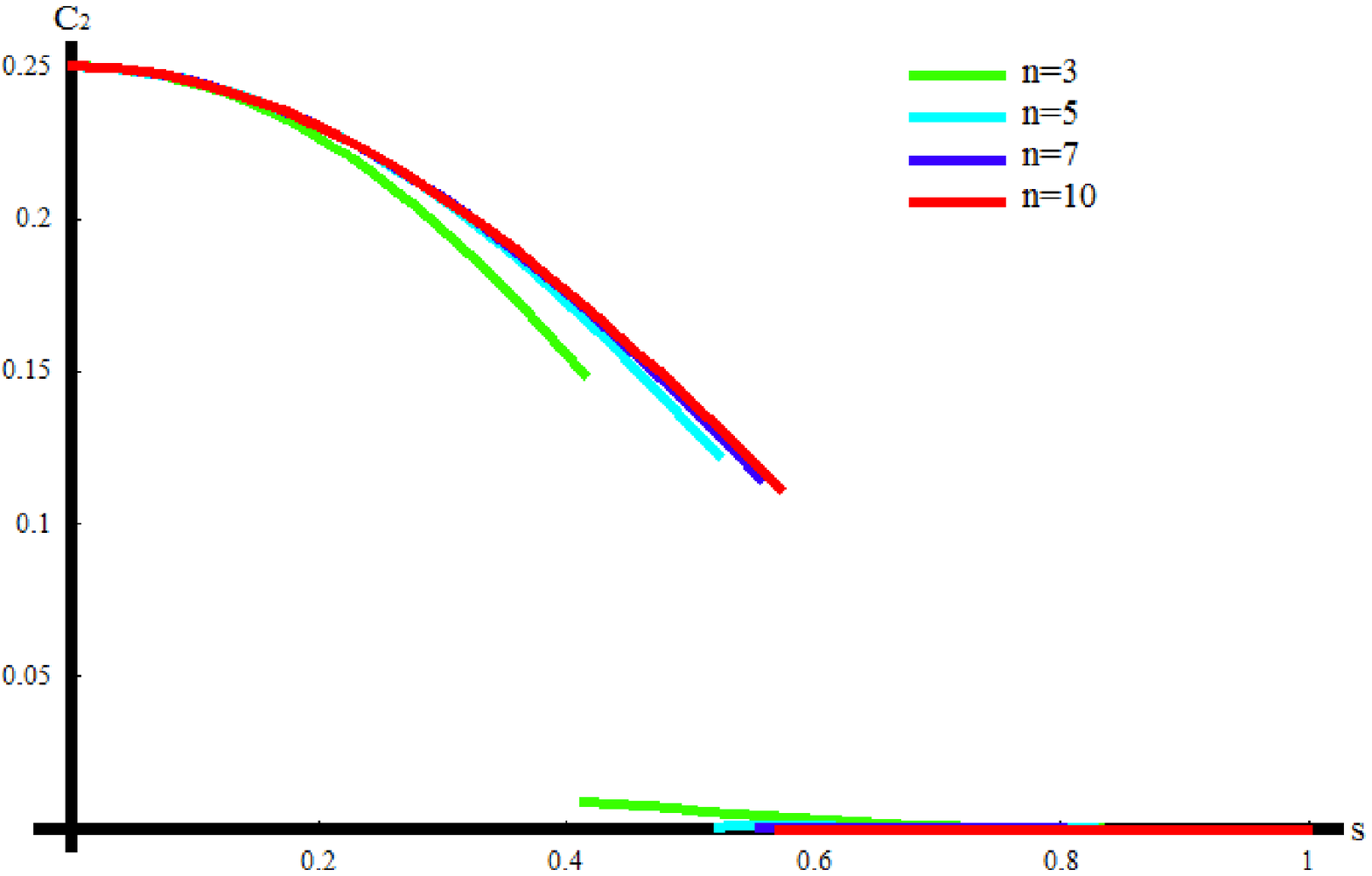}\\
{\bf Figure 3.}  {\sf The classical correlation $C_2$ versus the
overlapping $s$ for different values of $n$.}
\end{center}

\begin{center}
\includegraphics[width=4in]{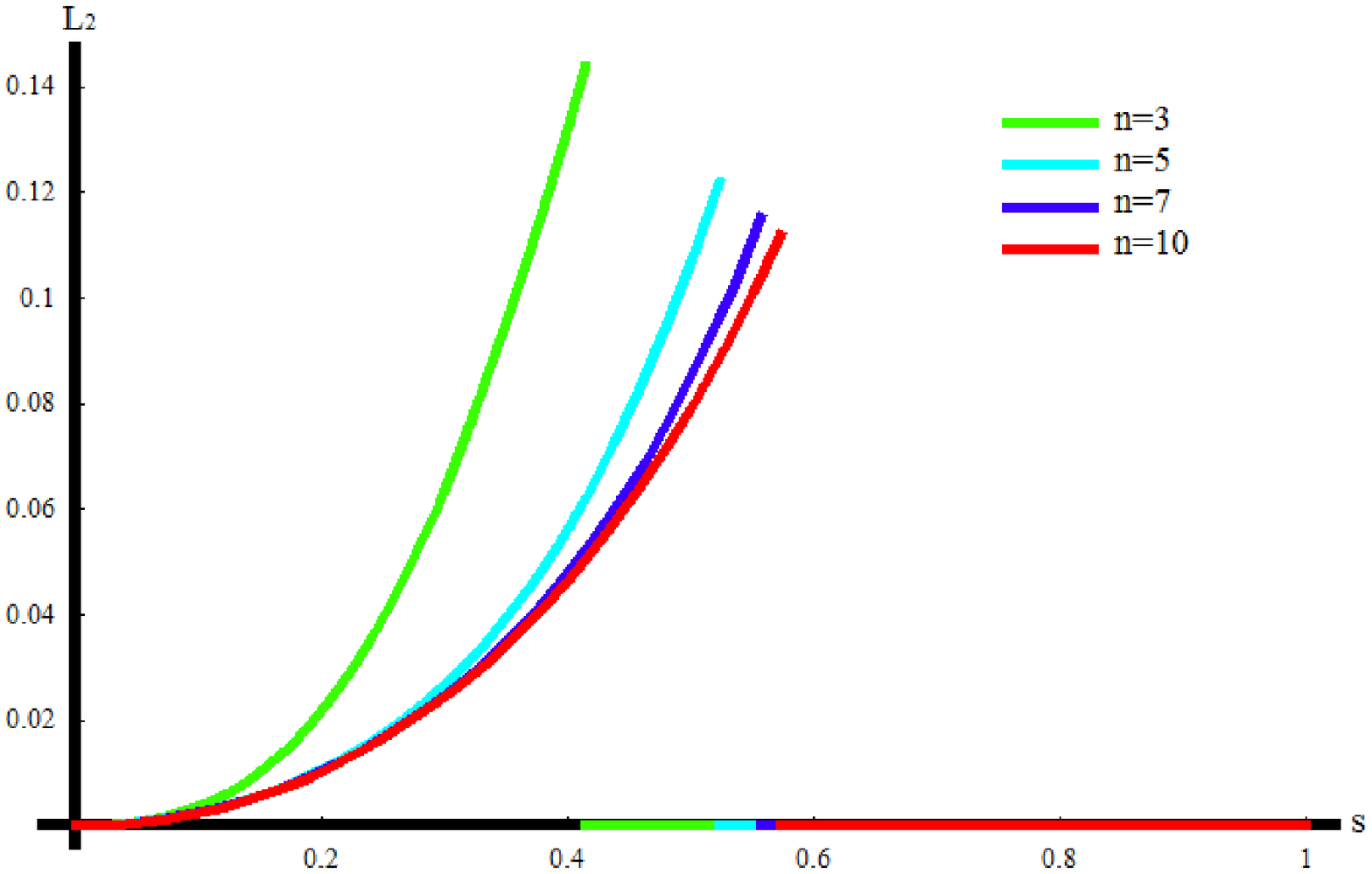}\\
{\bf Figure 4.}  {\sf The difference $L_2 = D_2+C_2-T_2$ versus the
overlapping $s$ for different values of $n$.}
\end{center}






\section{Concluding remarks}

In this paper, we have obtained explicit expressions for total,
classical and quantum pairwise correlations  using the
linear relative entropy. This variant of relative entropy, though
restricted from an informational point of view, constitutes an
advantageous tool to get analytically computable quantities. Furthermore, it
allows us to deal with the different kinds of correlations in a common
framework. This is reflected by the closed additivity relation among
the pairwise correlations. We deliberately considered a  $n$ qubit system, prepared in  states of the form (\ref{ncs}),
 to emphasize  the usefulness of
linear relative entropy in quantifying the correlations in a unified
way. To evaluate the amount of  the various pairwise correlations, two
different bi-partitioning schemes were examined. The first  scheme deals with
pure states and the second, obtained by a trace
procedure, concerns bipartite mixed  states. For bipartite pure  states, the total correlation
coincides with the sum of classical and quantum correlations.
However, in bipartite mixed  states, this relation is not always valid: $T_2 \neq D^+_2 + C^+_2$ (see equation (\ref{clos-add-rel})).
The non-vanishing quantity $L^+_2$ given by (\ref{L2+-}), measuring
the relative entropy between the product and product quantum-classical states, is
responsible of this inequality.  At this stage, it is interesting to
note that the sudden change of the quantum discord, when crossing the
transition point $\lambda_1 = \lambda_3$ (cf. (\ref{conditionplus})
and (\ref{conditionmoins})),  is accompanied by the apparition of the
non-vanishing quantity $L^+_2$ which diminishes the sum of classical
and quantum correlations in comparison with total correlation.
Finally, we notice that the obtained pairwise correlations are useful to
capture the essential of genuine multipartite correlation present in
the state (\ref{ncs}). The global multipartite correlation, defined as the
sum of all possible pairwise correlations in a  multi-qubit system (see for instance
\cite{Rulli,Ma}), can  be evaluated using the results derived here through the formalism of linear relative entropy.
We hope to report in this issue in a forthcoming work.

\end{document}